\documentclass{article}
\usepackage{enumerate}
\usepackage{amssymb}
\usepackage{amsthm}
\usepackage[intlimits,reqno]{amsmath}
\usepackage[mathcal]{eucal}
\usepackage{enumerate}
\usepackage{verbatim}

\newtheorem{theorem}{Theorem}[section]
\newtheorem{corollary}[theorem]{Corollary}
\newtheorem{lemma}[theorem]{Lemma}
\newtheorem{proposition}[theorem]{Proposition}
\theoremstyle{definition}
\newtheorem{definition}[theorem]{Definition}
\newtheorem{example}{Example}

\theoremstyle{remark}

\newcommand{\norm}[1]{\left\Vert#1\right\Vert}
\renewcommand{\sc}[2]{\langle #1|#2 \rangle}

\newcommand{\mn}[1]{\langle #1 \rangle}

\newcommand{\cat}[1]{| #1 \rangle}
\newcommand{\abs}[1]{\left\vert#1\right\vert}

\newcommand{\R}{\mathbb R}
\newcommand{\D}{\mathbb D}

\newcommand{\Z}{\mathbb Z}
\newcommand{\C}{\mathbb C}
\newcommand{\N}{\mathbb N}
\newcommand{\I}{\mathbb I}
\renewcommand{\P}{\mathcal{P}}
\newcommand{\CP}{\mathbb{CP}}

\newcommand{\A}{\mathcal{A}}
\newcommand{\B}{\mathcal{B}}
\newcommand{\K}{\mathcal{K}}
\renewcommand{\H}{\mathcal{H}}

\newcommand{\Li}{L^\infty(\mathcal{H})}

\renewcommand{\O}{\mathcal{O}}
\newcommand{\M}{\mathfrak{M}}

\newcommand{\pb}{\{\cdot,\cdot\}}

\renewcommand{\to}{\rightarrow}
\newcommand{\tto}{\longrightarrow}

\renewcommand{\leq}{\leqslant}
\renewcommand{\geq}{\geqslant}
\renewcommand{\phi}{\varphi}

\newcommand{\To}{\longrightarrow}

\DeclareMathOperator{\id}{id}

\DeclareMathOperator{\im}{im}

\newcommand{\be}{\begin{equation}}
\newcommand{\ee}{\end{equation}}
\newcommand{\bse}{\begin{subequations}}
\newcommand{\ese}{\end{subequations}}
\newcommand{\ben}{\begin{enumerate}}
\newcommand{\een}{\end{enumerate}}
\newcommand{\bit}{\begin{itemize}}
\newcommand{\eit}{\end{itemize}}

\renewcommand{\o}{\{0\}}
\newcommand{\E}{\mathbb{E}}
\renewcommand{\c}{\mathcal C}
\renewcommand{\M}{\mathcal{M}}
\renewcommand{\d}{\mathcal{D}}
\renewcommand{\L}{\mathbb L}
\renewcommand{\Li}{\pounds}

\DeclareMathOperator{\curv}{curv}
\DeclareMathOperator{\End}{End}

\usepackage[cp1250]{inputenc}

\begin{document}

\begin{center}
{\bf \Large Noncommutative K\"{a}hler-like structures in quantization\footnote{This work is supported in part by KBN grant 2 PO3 A 012 19.}}\\
\bigskip
{\large Anatol Odzijewicz  }\\ \bigskip
 Institute of Mathematics\\
 University  in Białystok\\
 ul. Lipowa 41, PL-15-424 Białystok, Poland\\
 e-mail: aodzijew@uwb.edu.pl
\vspace{1cm}
\end{center}

\begin{abstract}
One introduces the notion of $C^*$-algebra with polarization which could be considered as the
quantum K\"ahler structure. The connection of these algebras with Kostant-Souriau geometric
quantization is shown. The theory of polarized $C^*$-algebra is investigated by the use of the
coherent states method.
\end{abstract}

\tableofcontents

\section{Introduction}
There are complementary methods of mathematical description od quantum physical systems. These are the algebraic
methods based on the theory of $C^*$-algebras, see \cite{Emch}, and the geometric methods that find an elegant
presentation as Kostant-Souriau quantization and $*-$product quantization, see \cite{kostant2,souriau}.

In our approach we take an effort to unify both these methods by use of the notion of coherent states
map \cite{Ocoh}. The coherent states map $\K$ means a symplectic map of the classical phase space $M$
into quantum phase space, i.e. complex projective Hilbert space $\CP(\M)$. It appears that using
the
coherent states map one can unify, in some sense, the classical and quantum description of the
considered physical system. Furthermore the system is defined by $\K$ and Hamiltonian satisfying
some consistency condition with $\K$, see \cite{Oker,Ocoh,O-ber}.

Bearing in mind the above fact we introduce in Section \ref{4.0} the notion of a polarized algebra of observables $\A$,
which is univocally determined by coherent states map $\K$. In the case when $\K$ is Gaussian coherent states map of
linear phase space $\R^{2N}$ into $\CP(\M)$, the algebra $\A$ is Heisenberg-Weyl algebra. So the $C^*$-algebra $\A$ is
a natural generalization of the latter one to the case of a general phase space $M$. We prove some important properties
of $\A$ and explain the relation of the structure of $\A$ to the structures such as e.g. prequantum line bundle and
polarization which play a crucial role in Kostant-Souriau quantization.

As a result one can distinguish additional structures on $C^*$-algebra $\A$ that are responsible for the K\"ahler
structure of classical phase space $M$. This structure denotes the existence of a
commutative Banach subalgebra $\overline\P$ of $\A$ which has physical interpretation as the algebra of annihilation
operators. Its classical counterpart is the K\"ahler polarization in the sense of Kostant-Souriau quantization. So, it is
natural to understand $\overline\P$ as the quantum polarization and call $(\A,\overline\P)$ the polarized
$C^*$-algebras or quantum K\"ahler manifold.

In Section \ref{4} we introduce the notion of an abstract coherent state on $(\A,\overline\P)$. The
coherent states in this sense generalize the notion of vacuum to the case of a general phase space.
On the another hand using the coherent states
one can study the algebra $\A$ by reducing many problems to the investigation of its polarization
$\overline\P$, which is more handy because of commutativity.

Also in Section \ref{4} we show the fundamental properties of coherent states: on $\A$ they can be considered
as classical states of some classical phase space being subspace of space of multiplicative
functional on the polarization $\overline\P$. Additionally, when we apply the GNS construction to the
coherent states we obtain Hilbert space which is a generalization of Hardy space \cite{douglas,Rud}, which is exactly
obtained when $M=\D$ is a unit disc in $\C$ and $\A$ is Toeplitz algebra \cite{douglas}.

In Section \ref{5} we show how to reconstruct from $(\A,\overline\P)$ the classical phase space
$M$ and coherent states map $\K$. This gives rise to the method of reconstruction of the classical
mechanics picture for the quantum one.

Finally we would like to note that the results presented in the paper in fact is a step
in direction of the theory of
physical systems which unifies their classical and quantum description.

\section{Coherent state map and polarization}\label{2}
Let us fix a map $\K: M\to\CP(\M)$ of the manifold $M$ into
complex projective Hilbert space $\CP(\M)$. Assume that $\K$ is of
the same smoothness class as $M$ and that the image $\K(M)$ is
linearly dense in the Hilbert space $\CP(\M)$. This space will be
always assumed to be separable. For the
description of the map $\K$ as well as all related
objects one needs to fix a local trivialization
\be \label{eq:triv} K_\alpha: \Omega_\alpha\To\M\setminus\o \ee
 where
$[K_\alpha(m)]=\K(m)$ for $m\in\Omega_\alpha$ and \be
K_\alpha(m)=g_{\alpha\beta}(m)K_\beta(m) \ee for
$m\in\Omega_\alpha\cap\Omega_\beta$ of the map $\K$. The
open sets $\Omega_\alpha$, $\alpha\in J$, cover  the manifold $M$.
The transition functions
$$g_{\alpha\beta}:\Omega_\alpha\cap\Omega_\beta\To\C\setminus\o$$
form a smooth cocycle:
\be g_{\alpha\gamma}(m)g_{\gamma\beta}(m)=g_{\alpha\beta}(m),\qquad m\in\Omega_\alpha\cap\Omega_\beta\cap\Omega_\gamma \ee
on the manifold $M$.

Let $\mathbb L\to M$ denote the complex line bundle over $M$ being the
pull-back $\L:=\K^*\E$ of the universal bundle: \be
\E=\big\{(v,l)\in \M\times\CP(\M) : v\in l\big\}
\xrightarrow{pr_2} \CP(\M) \ee From the definition above it
follows that $\E\to\CP(\M)$ is the holomorphic line bundle
equipped with Hermitian metric $H_{FS}$ which is defined by the
scalar product $\langle\cdot|\cdot\rangle $ in the Hilbert space $\M$. Both
these structures uniquely determine the compatible connection
$\nabla_{FS}$ (see e.g. \cite{griffiths}). Since $i
\curv(\nabla_{FS})=:\omega_{FS}$ is the Fubini-Study form it is
natural to call
\be \big(\E\to\CP(\M),\nabla_{FS},H_{FS},\omega_{FS}\big) \ee
the Fubini-Study pre-quantum bundle. According to \cite{NarRam} it is the
universal object of the category of pre-quantum bundles in the sense
of B. Kostant \cite{kostant2,souriau}. The above means that for any pre-quantum bundle
\be \label{eq:bun} \big(\L\to M,\nabla,H,\omega\big) \ee
there exists a map $\K:M\to\CP(\M)$ such that \eqref{eq:bun} is given as
the pull-back of the Fubini-Study bundle: \be \L=\K^*\E,\quad
\nabla=\K^*\nabla_{FS},\quad H=\K^*H_{FS},\quad
\omega=\K^*\omega_{FS}. \ee

Let us recall that objects presented in \eqref{eq:bun} satisfy appropriate consistency conditions and
the condition $[\omega]\in H^2(M,\Z)$.

In the terms of trivialization \eqref{eq:triv} the basic sections $s_\alpha:\Omega_\alpha\to\L$ are given by
\be s_\alpha(m):=\big(m,K_\alpha(m)\big),\quad m\in\Omega_\alpha \ee
and the related potentials of the Hermitian metric H are
\be H(s_\alpha,s_\alpha)=\langle K_\alpha|K_\alpha\rangle . \ee
While the connection form
\be \nabla s_\alpha=:\Theta_\alpha\otimes s_\alpha \ee
can be written as
\be \Theta_\alpha=\frac{\langle K_\alpha|dK_\alpha\rangle }{\langle K_\alpha|K_\alpha\rangle } \ee
For the sake of completeness we give the expression for the curvature 2-form
\be \omega=i \curv\nabla=i \; d\frac{\langle K_\alpha|dK_\alpha\rangle }{\langle K_\alpha|K_\alpha\rangle }\ee

These formulae will be useful in the sequel.

Let $\overline{\L}^*\to M$ be the line bundle dual to the complex conjugate $\overline\L\to M$ of $\L\to M$. This bundle is also equipped with the metric
structure $\overline H^*$ and the connection $\overline\nabla^*$. Their expressions in the gauge $\overline s^*_\alpha:\Omega_\alpha\to\overline\L^*$ are
given by:
\be \overline H^*(\overline s_\alpha^*,\overline s_\alpha^*)=\frac{1}{\langle K_\alpha|K_\alpha\rangle } \ee

\be \label{eq:theta}\overline\Theta_\alpha^*=\frac{\langle dK_\alpha|K_\alpha\rangle }{\langle K_\alpha|K_\alpha\rangle } \ee
where $\overline\nabla^*\overline s_\alpha^*=\overline\Theta_\alpha^*\otimes\overline s_\alpha^*$ on $\Omega_\alpha$.

Let us consider now the map:
\be \label{eq:I} I(v)(m):=\langle K_\alpha(m)|v\rangle \overline s_\alpha^*(m),\quad m\in\Omega_\alpha. \ee
of the Hilbert space $\M$ into the vector space $\Gamma(M,\overline\L^*)$ of the smooth sections of the bundle $\overline\L^*$. It is easy to check that the
definition above does not depend on the choice of the gauge. The map $I$ is a linear injection of $\M$ into $\Gamma(M,\overline\L^*)$. So the Hilbert space $\M$
can be identified with the vector subspace $I(M)\subset \Gamma(M,\overline\L^*)$.

Let us consider the complex distribution $P\subset T^\C M$
spanned by smooth complex
vector fields $X\in\Gamma^\infty(T^\C M)$
which annihilate the Hilbert space $I(\M)\subset \Gamma^\infty(M,\overline\L^*)$, i.e.
\be P:=\bigsqcup_{m\in
M} P_m,\ee
where \be P_m:=\{X(m): X\in\Gamma^\infty(T^\C M) \textrm{ and } \overline\nabla^*_X\psi=0 \textrm{ for any } \psi\in I(\M)\}.\ee

To summarize the properties of $P$ we formulate
\begin{proposition} \label{prop:5}
$\quad$
\begin{enumerate}[i)]
\item The necessary and sufficient condition for $X$ to belong to
$\Gamma^\infty(P)$ is
\be \label{PX}\overline X(K_\alpha)=\Theta_\alpha(\overline
X)K_\alpha. \ee
\item The distribution $P$ is involutive and isotropic, i.e. for
$X,Y\in\Gamma^\infty(P)$ one has
\be [X,Y]\in\Gamma^\infty(P)\quad\textrm{and }\omega(X,Y)=0.\ee
\item If $X\in\Gamma^\infty(P\cap\overline P)$ then
\be \label{eq:zero}X\llcorner\omega=0\ee
\item Positivity condition:
\be \label{eq:posit} i \omega(X,\overline X)\geq 0 \ee
 for all $X\in\Gamma^\infty(P)$.
\end{enumerate}
\end{proposition}
\begin{proof}
$\quad$
\begin{enumerate}[i)]
\item By the definition one has that $X\in\Gamma^\infty(P)$ iff
$\overline\nabla^*_X I(v)=0$ for any $v\in\M$. From \eqref{eq:theta}
and \eqref{eq:I} we get
$$\overline\nabla^*_XI(v)=X\llcorner\left(\langle dK_\alpha|v\rangle -\frac{\langle dK_\alpha|K_\alpha\rangle }{\langle K_\alpha|K_\alpha\rangle }\langle K_\alpha|v\rangle \right)
\otimes\overline s_\alpha^*=$$
$$=\langle \overline
X(K_\alpha)|\left(\I-\frac{|K_\alpha\rangle \langle K_\alpha|}{\langle K_\alpha|K_\alpha\rangle }\right)v>\overline
s_\alpha^*=\langle \overline X(K_\alpha)-\Theta_\alpha(\overline X)K_\alpha|v>\overline s^*_\alpha$$
Thus we have proven \eqref{PX}.

\item From
\be d\Theta_\alpha=d\frac{\langle K_\alpha|dK_\alpha\rangle }{\langle K_\alpha|K_\alpha\rangle }=
\frac{\langle dK_\alpha\wedge|dK_\alpha\rangle }{\langle K_\alpha|K_\alpha\rangle }-\frac{\langle dK_\alpha|K_\alpha\rangle \wedge \langle K_\alpha|dK_\alpha\rangle }{(\langle K_\alpha|K_\alpha\rangle )^2}\ee
and \eqref{PX} we obtain
$$d\Theta_\alpha(X,Y)=\frac12\left(\frac{\langle \overline
X(K_\alpha)|Y(K_\alpha)\rangle -\langle \overline
Y(K_\alpha)|X(K_\alpha)\rangle }{\langle K_\alpha|K_\alpha\rangle }-\right.$$
$$\left.-\frac{\langle \overline
X(K_\alpha)|K_\alpha\rangle \langle K_\alpha|Y(K_\alpha)\rangle -\langle \overline
Y(K_\alpha)|K_\alpha\rangle  \langle K_\alpha|X(K_\alpha)\rangle }{(\langle K_\alpha|K_\alpha\rangle )^2}\right)=$$
\be \label{eq:dtheta}=\frac12\left(
\overline{\Theta_\alpha(\overline
X)}\Theta_\alpha(Y)-\overline{\Theta_\alpha(\overline
Y)}\Theta_\alpha(X)-\overline{\Theta_\alpha(\overline
X)}\Theta_\alpha(Y)+\overline{\Theta_\alpha(\overline
Y)}\Theta_\alpha(X)\right)=0 \ee
for $X,Y\in\Gamma^\infty(P)$.
Using the identity
\be
d\Theta_\alpha(X,Y)=\nabla_X\nabla_Y-\nabla_Y\nabla_X-\nabla{[X,Y]}\ee
and \eqref{eq:dtheta} we conclude that $P$ is involutive isotropic
distribution.
\item Let $X\in\Gamma^\infty(P\cap\overline P)$, then
$$\Li_X\Theta_\alpha=\Li_X\frac{\langle K_\alpha|dK_\alpha\rangle }{\langle K_\alpha|K_\alpha\rangle }=-\frac{1}{\langle K_\alpha|K_\alpha\rangle }\left(
\langle \overline
X(K_\alpha)|K_\alpha\rangle +\right.$$
$$+\left.\langle K_\alpha|X(K_\alpha)\rangle \right)\Theta_\alpha+\frac{1}{\langle K_\alpha|K_\alpha\rangle }\left(\langle \overline
X(K_\alpha)|dK_\alpha\rangle +\langle K_\alpha|d
X(K_\alpha)\rangle \right)=$$
$$=-\left(\overline{\Theta_\alpha(\overline
X)}+\Theta_\alpha(X)\right)\Theta_\alpha+\overline{\Theta_\alpha(\overline
X)}\Theta_\alpha+\Theta_\alpha(X)\Theta_\alpha+d\left(\Theta_\alpha(X)\right)=$$
\be d\left(\Theta_\alpha(X)\right)+
X\llcorner d\Theta_\alpha-X\llcorner
d\Theta_\alpha=\Li_X\Theta_\alpha-X\llcorner d \Theta_\alpha \ee
Hence one has \eqref{eq:zero}.
\item For $X\in\Gamma^\infty(P)$ one can write
$$d\Theta_\alpha(X,\overline X)=\frac12\left(\overline{\Theta_\alpha(\overline X)}\Theta_\alpha(\overline X)-\right.$$
$$\left.-\frac{\langle X(K_\alpha)|X(K_\alpha)\rangle }{\langle K_\alpha|K_\alpha\rangle }-\overline{\Theta_\alpha(\overline X)}\Theta_\alpha(\overline X)
+\frac{\langle X(K_\alpha)|K_\alpha\rangle \langle K_\alpha|X(K_\alpha)\rangle }{(\langle K_\alpha|K_\alpha\rangle )^2}\right)=$$
\be \label{eq:32} =-\frac{1}{2\norm{K_\alpha}^2}\left(\norm{X(K_\alpha)}^2\norm{K_\alpha}^2-\abs{\langle K_\alpha|X(K_\alpha)\rangle }^2\right) .\ee
Now from Schwartz inequality one gets \eqref{eq:dtheta}.
\end{enumerate}
\flushright\qed\end{proof}

\begin{definition}
Let $\mathcal O_\mathcal K$ denote the algebra of functions $\lambda\in C^\infty(M)$ such that $\lambda\psi\in I(\M)$ if $\psi\in I(M)$.
\end{definition}

In all further considerations we shall restrict ourselves to the cases of maps $\K:M\To\CP(\M)$ which do satisfy the following conditions:
{\it \begin{enumerate}[a)]
\item The curvature $2$-form
$$\omega=i\curv\nabla=\K^*\omega_{FS}$$
is non-degenerate, i.e. $\omega$ is symplectic.
\item The distribution $P$ is maximal. i.e.
\be \dim_\C P=\frac{1}{2}\dim M=:N.\ee
\item For every $m\in M$ there exists an open neighborhood $\Omega\ni m$ and functions $\lambda_1,\ldots,\lambda_N\in \O_\K$ such that $d\lambda_1,\ldots,d\lambda_N$ are
linearly independent on $\Omega$.
\end{enumerate}}

According to \cite{Ocoh} the map $\K$ satisfying given conditions will be called a {\bf coherent state map}.
We present the following properties of the map $\K$.
\begin{proposition}
$\quad$
\begin{enumerate}[i)]
\item The manifold $M$ is K\"ahler manifold and $\K:M\To \CP(\M)$ is a K\"ahler immersion of $M$ into $\CP(\M)$.
\item The distribution $P$ is K\"ahler polarization of symplectic manifold $(M,\omega)$. Moreover $P$ is spanned by the Hamiltonian vector fields $X_\lambda$ generated
by $\lambda\in\O_\K$.
\end{enumerate}
\end{proposition}
\begin{proof}
$\quad$
\begin{enumerate}[i)]
\item The condition $a)$ and $b)$ imply that $P$ define almost complex structure on $M$. The condition $c)$ guarantees its integrability. The property of M being K\"ahler
follows from the fact that $\omega$ is symplectic and from positivity property \eqref{eq:posit} of Proposition \ref{prop:5}. The immersion property
of $\K$ follows from $\omega=\K^*\omega_{FS}$ and $\omega$ is symplectic.
\item Let us take $X\in\Gamma^\infty(P)$ and $\lambda\in\O_\K$. Then from
$$\overline\nabla^*_X\psi=0\qquad\textrm{and}\qquad \overline\nabla^*_X(\lambda\psi)=0$$
for any $\psi\in I(\M)$, it follows $X(\lambda)=0$. Let $X_\lambda$ be the Hamiltonian vector field corresponding to $\lambda$
\be X_\lambda\llcorner\omega=d\lambda \ee
Then
$$\omega(X_\lambda,X)=d\lambda(X)=X(\lambda)=0.$$
Since $P$ is maximal isotropic one gets $X_\lambda\in\Gamma^\infty(P)$. The condition $c)$ implies now that $P$ is spanned by $X_\lambda$ where $\lambda\in\O_\K$.
In this way we have shown that $P$ is integrable K\"ahler polarization on $(M,\omega)$.
\end{enumerate}
\flushright\qed\end{proof}

We conclude this section by making the following comment. In the symplectic case the Lie subalgebra $(\O_\K,\{\cdot,\cdot\})$
is  a maximal commutative subalgebra of the algebra of classical observables $(C^\infty(M),\pb)$. The corresponding
Hamiltonian vector fields $X_{\lambda_1},\ldots X_{\lambda_N}\in \Gamma^\infty(P)$ span the {\bf K\"ahler polarization}
P in the sense of Kostant-Souriau geometric quantization.

\section{Quantum polarization}\label{3}
Let $\d$ be the vector subspace of the Hilbert space $\M$ generated by finite combinations of the vectors $K_\alpha(m)$, where $\alpha\in I$ and $m\in\Omega_\alpha$.
The linear operator $a:\d\to \M$ such that
\be \label{eq:anih}aK_\alpha(m)=\lambda(m)K_\alpha(m) \ee
for any $\alpha\in I$ and $m\in\Omega_\alpha$, will be called the {\bf annihilation operator}. On the other hand the operator $a^*$
conjugated to $a$ we shall call the {\bf creation operator}. The eigenvalue function $\lambda:M\to\C$ is well
defined on $M$ since $K_\alpha(m)\neq 0$ and the condition \eqref{eq:anih} does not depend on the choice of gauge.

In general the annihilation operators are not bounded as it is in the case of the Gaussian coherent states map
(see Example \ref{q-heis} of Section \ref{4.0}). In this paper we restrict ourselves
to the case when the annihilation operators are bounded.
\begin{proposition}
The bounded annihilation operators form a commutative unital Banach subalgebra $\overline\P_{\mathcal K}$ in the algebra $\B(\M)$ of all bounded operators in the Hilbert space $\M$.
\end{proposition}
\begin{proof}
It follows directly from the definition \eqref{eq:anih} that for any elements $a_1,a_2\in\overline\P_{\mathcal K}$ their product $a_1a_2$ and linear combination $c_1a_1+c_2a_2$ belong
to $\overline\P_{\mathcal K}$. It is also clear that identity operator $\I\in\overline\P_{\mathcal K}$. We shall show completeness of the subalgebra $\overline\P_{\mathcal K}\subset\B(\M)$. Let $\{a_n\}_{n\in\N}$
 be a Cauchy sequence of annihilation operators and let $\{\lambda_n\}_{n\in\N}$ be the corresponding sequence of their eigenfunctions. From the condition
 \eqref{eq:anih}
 $$\abs{\lambda_k(m)-\lambda_n(m)}=\frac{\norm{(a_k-a_n)K_\alpha(m)}}{\norm{K_\alpha(m)}}\leq\norm{a_k-a_n} $$
for all $\alpha\in I$ and $m\in\Omega_\alpha$. Hence the sequence $\{\lambda_k\}_{k\in\N}$ converges pointwise to some function
$\lambda:\M\to\C$. Since $a_n\xrightarrow{n\to\infty}a$
converges in the operator norm to some bounded operator $a\in\B(\M)$ one has
$$\norm{(a-\lambda(m)\I)K_\alpha(m)}=\lim_{n\to\infty}\norm{(a_n-\lambda_k(m)K_\alpha(m)}=0 $$
Consequently $\lambda$ is the eigenvalues function for $a$ and $a\in\overline\P_{\mathcal K}$. The annihilation
operators $a_1$, $a_2\in \overline\P_{\mathcal K}$ commute on a dense domain $\d\subset\M$ implying the commutativity
of the subalgebra $\overline\P_{\mathcal K}$. \flushright\qed\end{proof}

The eigenvalues function is the covariant symbol \be \label{eq:mean}\lambda(m)=\frac{\langle K_\alpha(m)|a
K_\alpha(m)\rangle }{\langle K_\alpha(m)|K_\alpha(m)\rangle }=:\langle a\rangle (m) \ee of the annihilation operator. It is thus a bounded complex
analytic function on the complex manifold $M$.

We shall describe now the algebra $\mathcal P_\K:=\{a^*, a\in \overline{\mathcal P}_\K\}$ of creation operators
in terms of the Hilbert space $I(\M)$.
 Let $\Lambda:\M\to\M$ be the linear operator defined in the following way:
there exists $\lambda\in\mathcal O_\K$ such that
$$\lambda I(v)=I(\Lambda v)$$
for all $v\in\M$. The operator defined above has the following properties.
\begin{proposition}\label{pr:lambda}
$\quad$
\begin{enumerate}[i)]
\item The operator $\Lambda$ is bounded on $\M$.
\item The operator  $\Lambda^*$ adjoint to $\Lambda$ is an annihilation operator with the covariant symbol given by
bounded function $\overline\lambda$.
\end{enumerate}
\end{proposition}
\begin{proof}
$\quad$
\begin{enumerate}[i)]
\item From the sequence of equalities
$$\langle \overline{\lambda(m)}K_\alpha(m)|v\rangle \overline s^*_\alpha(m)=\lambda(m)\langle K_\alpha(m)|v\rangle \overline
s^*_\alpha=$$
\be \label{eq:lambda}=\langle K_\alpha(m)|\Lambda v\rangle
\overline s^*_\alpha(m) \ee
where $v\in\M$, $\alpha\in J$ and $m\in\Omega_\alpha$ it follows that $\d$ is the domain of the conjugated operator $\Lambda^*$. Since
$\d$ is dense in $\M$ the operator $\Lambda$ admits the closure $\overline\Lambda=\Lambda^{**}$, see \cite{akhiezer}. We have $\M=D(\Lambda)\subset D(\overline\Lambda)$
which implies the boundedness of $\Lambda$.
\item Let us notice that from \eqref{eq:lambda} it follows
\be \Lambda^*K_\alpha(m)=\overline\lambda(m)K_\alpha(m). \ee
Thus $\Lambda^*$ is the annihilation operator with $\overline\lambda$ as its covariant symbol.
\end{enumerate}
\flushright\qed\end{proof}

From this two propositions above one can deduce the following.
\begin{proposition}
The mean value map $\mn\cdot$ defined by \eqref{eq:mean} gives the continuous
\be \label{eq:meannorm}\norm{\mn b}_\infty=\sup_{m\in M}\abs{\mn{b}(m)}\leq\norm b \ee
isomorphism of $\overline{\P_{\mathcal K}}$ with the function Banach algebra $(\O_\K,\norm{\cdot}_\infty)$.
\end{proposition}

Let us assume that for some measure $\mu$ one has the resolution of the identity operator
\be \label{eq:iden} \I=\int_M P(m)d\mu(m), \ee
where
\be \label{eq:proj} P(m):=\frac{|K_\alpha(m)\rangle \langle K_\alpha(m)|}{\langle K_\alpha(m)|K_\alpha(m)\rangle } \ee
is the operator of orthogonal projection on the coherent state $\K(m)$, $m\in M$.
In such case the scalar product of the functions $\psi=I(v)$ and $\phi=I(w)$ can be expressed in terms of the integral
$$\sc\psi\phi=\sc vw=\int_M \overline{H}^*(\psi,\phi)d\mu=$$
\be \label{eq:scal}=\int_M\frac{\overline{\langle K_\alpha(m)|v\rangle }\langle K_\alpha(m)|w\rangle }
{\langle K_\alpha(m)|K_\alpha(m)\rangle }d\mu(m). \ee
Moreover one has
\be\norm{\Lambda v}^2=\int_M \abs\lambda^2\overline H^*(\psi,\psi)d\mu\leq\norm\lambda_\infty^2\norm v^2 \ee
for $v\in\M$ and thus it follows that
\be \label{eq:lambdanorm}\norm \Lambda\leq\norm\lambda_\infty. \ee

Taking into account the inequalities \eqref{eq:meannorm} and \eqref{eq:lambdanorm} we obtain
\begin{theorem} \label{thm:4}
If the coherent states map admits the measure $\mu$ defining the resolution  of
identity \eqref{eq:iden} then the mean value map
$\mn\cdot$ is the isometric isomorphism of the Banach algebra
$(\overline\P_{\mathcal K},\norm\cdot)$ onto Banach algebra
$(\O_\K,\norm\cdot_\infty)$.
\end{theorem}
From the theorem above one may draw the conclusion that the
necessary condition for the existence of the identity
decomposition for the coherent states map $\K$ is the uniformity
of the algebra $\overline\P_{\mathcal K}$, i.e.
$$\norm{a^2}=\norm a^2\quad\textrm{for }a\in\overline\P_{\mathcal K}.$$

We shall show some facts allowing better understanding of the
covariant symbols algebra $\O_\K$ in the context of the geometric
quantization.
According to $ii)$ of Proposition \ref{pr:lambda} it is easy to notice that Kostant-Souriau quantization
\be \O_\K\ni\lambda\To Q_\lambda=i\nabla_{X_\lambda}+\lambda \ee
gives the realization of $\P_\K$ in the Hilbert space $I(M)\subset \Gamma^\infty(M,\overline{\mathbb L}^*)$.

In the light of the remarks above it is strongly justified to call the Banach algebra $\overline\P_{\mathcal K}$ a {\bf quantum K\"ahler polarization}
of the mechanical
system defined by K\"ahler coherent immersion $\K:M\To\CP(\M)$.

The next section will be dedicated to purely quantum description of the mechanical system in $C^*$-algebra approach.
\section{Polarized $C^*$-algebras as a quantum K\"ahler phase spaces}\label{4.0}

The function algebra $\O_\K$  defines the complex analytic coordinates of the classical phase space $(M,\omega)$, i.e. for any
$m\in M$ there are open neighborhoods $\Omega\ni m_0$ and $z_1,\ldots,z_N\in\O_\K$ such that the map $\phi:\Omega\to \C^N$ defined
by $\phi(m):=(z_1(m),\ldots,z_N(m))$ for $m\in\Omega$, is a holomorphic chart from the complex analytic atlas of $M$. The
{annihilation operators} $a_1,\ldots,a_N\in \overline\P_{\mathcal K}$ correspond to $z_1,\ldots,z_N$ through the defining relation \eqref{eq:anih}
is naturally to consider as a quantum complex coordinate system.

Let us define {\bf Berezin covariant symbol}
\be \mn F(m)=\frac{\sc{K_\alpha(m)}{FK_\alpha(m)}}{\sc{K_\alpha(m)}{K_\alpha(m)}},\qquad m\in M\ee
of the operator $F$ (unbounded in general) which domain $\mathcal D$ contains all finite linear combinations
of coherent states. Since $\K:M\to \CP(\M)$ is a complex analytic map, the Berezin covariant symbol $\mn F$ is a real analytic
function  of the coordinates $\overline z_1,\ldots,\overline z_N,z_1,\ldots,z_N$.

For $n\in\N$ let $F_n(a_1^*,\ldots,a_N^*,a_1,\ldots,a_N)\in A_\K$
be a normally ordered polynomials of creation and annihilation operators. We say that
\be F_n(a_1^*,\ldots,a_N^*,a_1,\ldots,a_N)\xrightarrow[n\to\infty]{} F=:F(a_1^*,\ldots,a_N^*,a_1,\ldots,a_N)\ee
converges in {\bf coherent state weak topology} if
\be \mn{F_n(a_1^*,\ldots,a_N^*,a_1,\ldots,a_N)}(m)\xrightarrow[n\to\infty]{} \mn F(m).\ee

Therefore thinking about observables of the considered system,
i.e. self-adjoint operators, as the weak coherent state limits of normally ordered polynomials of annihilation and creation operators
we are justified to assume the following
\begin{definition}
  The unital operator $C^*$-algebra $\A_\K$ generated by the Banach algebra $\overline\P_\K$ we shall call {\bf quantum K\"ahler phase space}
  generated by the coherent state map $\K:M\to\CP(\H)$.
\end{definition}

Taking into account the properties of $\A_\K$ we define the abstract polarized $C^*$-algebra.
\begin{definition}
The {\bf polarized $C^*$-algebra} is a pair $(\A,\overline\P)$ consisting of the unital $C^*$-algebra $\A$ and its Banach commutative subalgebra $\overline\P$
such that
\begin{enumerate}[i)]
\item $\overline\P$ generates $\A$
\item $\overline\P\cap\P=\C\I$
\end{enumerate}
\end{definition}

It is easy to see that $\A_\K$ is polarized $C^*$-algebra in the sense of this definition.

Also the notion of coherent state can be generalized to the case of abstract polarized $C^*$-algebra $(\A, \overline\P)$, namely
\begin{definition} \label{defn:coh}
A {\bf coherent state} $\omega$ on polarized $C^*$-algebra $(\A,\overline\P)$ is the positive linear functional of the norm equal to one satisfying the condition
\be \label{eq:coh}\omega(xa)=\omega(x)\omega(a) \ee
for any $x\in\A$ and any $a\in\overline\P$.
\end{definition}
Let us stress that in the case when $(\A,\overline\P)$ is defined by the coherent state map $\K:M\to\CP(\M)$ then the state
\be \omega_m(x):=Tr\big(xP(m)\big), \ee
where $m\in M$ and $P(m)$ is given by \eqref{eq:proj}, is coherent in the sense of Definition \ref{defn:coh}

Proceeding as in motivating remarks we shall introduce the notion of the norm normal ordering in polarized $C^*$-algebra $(\A,\overline\P)$.
\begin{definition} The $C^*$-algebra $\A$ of quantum observables with fixed polarization $\overline\P$ admits the {\bf norm normal ordering} if and only if the set of elements
of the form
$$\sum_{k=1}^N b_k^*a_k$$
where $N\in\N$ and $a_1,\ldots,a_N,b_1,\ldots,b_N\in\overline\P$, is dense in $\A$ in $C^*$-algebra norm topology.
\end{definition}
Since we assume that $\A$ is unital the coherent states on $(\A,\overline\P)$ are positive continuous functionals satisfying the condition $\omega(\I)=1$. The set of all coherent states on
$(\A,\overline\P)$ will be denoted by $\c(\A,\overline\P)$. The structure of $\c(\A,\overline\P)$ is investigated and described in the next section of this paper. Some properties of
coherent states are however needed now for the description of algebra $\A_\K$ defined by the coherent state map $\K:M\to \CP(\M)$.
\begin{theorem}\label{thm:coh}
Let $\rho\neq 0$ be a positive linear functional on $(\A,\overline\P)$. Assume that $\rho\leq \omega$, where $\omega\in\c(\A,\overline\P)$ is a coherent state. Then
\begin{enumerate}[i)]
\item the functional $\frac{1}{\rho(\I)}\rho$ is the coherent state and
$$ \frac{1}{\rho(\I)}\rho(a)=\omega(a)$$
for $a\in\overline\P$.
\item If $(\A,\overline\P)$ admits the norm normal ordering then
$$ \frac{1}{\rho(\I)}\rho=\omega,$$
i.e. the coherent state $\omega$ is pure.
\end{enumerate}
\end{theorem}
\begin{proof}
$\quad$
\begin{enumerate}[i)]
\item Let $\pi_\omega:\A\To\H_\omega$ be the GNS representation of $\A$ and let $v_\omega\in\H_\omega$ be the cyclic vector of this reprezentation
corresponding to $\omega$. Then there exists an operator $T\in\pi_\omega(\A)'$, $0\leq T\leq 1$, such that
\be \label{eq:40}\rho(x)=\langle Tv_\omega|\pi_\omega(x)Tv_\omega\rangle  \ee
for any $x\in\A$, see \cite{dixmier,murphy}. From the defining property \eqref{eq:coh} of the coherent state one gets
$$\langle v_\omega|\pi_\omega(x)(\pi_\omega(a)-\omega(a))v_\omega\rangle =0.$$
Since $v_\omega$ is cyclic for $\pi_\omega(\A)$ it must be
\be \label{eq:41} \pi_\omega(a)v_\omega=\omega(a)v_\omega \ee
for any $a\in\overline\P$. From \eqref{eq:40} and \eqref{eq:41} it follows that
\be \label{eq:42} \rho(xa)=\rho(x)\omega(a) \ee
for an $x\in\A$ and $a\in\overline\P$. Taking $x=\I$ in \eqref{eq:42} we get $ \frac{1}{\rho(\I)}\rho(a)=\omega(a)$.
Substituting $\omega(a)=\frac{1}{\rho(\I)}\rho(a)$ into \eqref{eq:42} and dividing both sides of \eqref{eq:42} by $\rho(\I)\neq 0$
we find that $\frac{1}{\rho(\I)}\rho(a)$ belongs to $\c(\A,\overline\P)$.
\item Since $\frac{1}{\rho(\I)}\rho(a)$ is equal to $\omega$ on $\overline\P$ we have
$$\frac{1}{\rho(\I)}\rho\!\!\left(\sum_{k=1}^N b_k^*a_k\!\!\right)\!\!=\!\sum_{k=1}^N\overline{\frac{1}{\rho(\I)}\rho(b_k)}\frac{1}{\rho(\I)}\rho(a_k)=
\!\sum_{k=1}^N\overline{\omega(b_k)}\omega(a_k)=\omega\!\!\left(\sum_{k=1}^N b_k^*a_k\!\!\right).$$
From the existence of the normal ordering on $(\A,\overline\P)$ and continuity of $\rho$ and $\omega$ it follows that $ \frac{1}{\rho(\I)}\rho=\omega$ on $\A$.
\end{enumerate}
\flushright\qed\end{proof}

Let us remark that the norm normal ordering property of the polarized $C^*$-algebra $\A$ is stronger than the normal ordering in
the Heisenberg quantum mechanics or quantum field theory where it is considered in the weak topology sense.

One of the commonly accepted principles of
quantum theory is irreducibility of the algebra of quantum observables. For the Heisenberg-Weyl algebra case the irreducible representations are
equivalent to Schr\"odinger representation due to
the von Neumann theorem \cite{reed1}. In the case of general coherent states map $\K:M\to\CP(\M)$ the irreducibility of the corresponding algebra $\A_\K$ of
observables depends on the existence of the norm normal ordering.
\begin{theorem}\label{thm:irr}
Let $\A_\K$ be polarized algebra of observables defined by the coherent states map $\K:M\to\CP(\M)$. If $M$ is connected and
there exists the norm normal ordering on $\A_\K$ then the auto-representation $\id:\A_\K\To\B(\M)$ is irreducible.
\end{theorem}
\begin{proof}
It was stated in Theorem \ref{thm:coh} that the vector coherent state $\K(m_1)$ is pure one. This implies irreducibility of representation
$$\pi_{m_1}:=\id_{|\M_{m_1}}:\A_\K\To\End\M_{m_1}$$
of the algebra $\A_\K$ in the Hilbert subspace $\M_{m_1}=\A\K(m_1)$. There are two possibilities: either $\K(m_2)\subset\M_{m_1}$ for any
$m_2\in M$ or there exists $m_2\in M$ such that $\K(m_2)\nsubseteq\M_{m_1}$. In the second case it follows from irreducibility of representation
$$\pi_{m_2}:=\id_{|\M_{m_2}}:\A_\K\To\End\M_{m_2}$$
that $\M_{m_2}\subset\M_{m_1}^\bot$. Applying this procedure step by step one obtain the orthogonal decomposition
$$\M=\bigoplus_{i\in I} \M_{m_i}$$
of the Hilbert space $\M$. From assumed separability of $\M$ we find that $I$ is at most countable.

Let
$$M_i:=\big\{m\in M: \K(m)\subset\M_{m_i}\big\}$$
where $i\in I$. If $m\in M_i\cap\Omega_\alpha$ then
$$\langle K_\alpha(m)|K_\alpha(m)\rangle \;>0.$$
Since $K_\alpha:\Omega_\alpha\to\C$ is continuous there exists a open neighborhood $m\in\O\subset\Omega_\alpha$ of $m$ such that
$$\langle K_\alpha(m)|K_\alpha(m')\rangle \neq0$$
for $m'\in\O$. The following inclusion must be valid $\K(\O)=\big[K_\alpha(\O)\big]\subset\M_{m_i}$. Otherwise one would have
$$\langle K_\alpha(m)|K_\alpha(m')\rangle =0$$
which contradicts the definition of the set $\O$. In this way we have shown that $\O\subset M_{i}$ and $M_{i}$ is open in $M$. Thus
M is disjoint union
$$M=\bigcup_{i\in I} M_{i}$$
of the open sets. Since, by assumption, M is connected it must be $M=M_{i}$ for some $i\in I$. The above means that $\M=\A_\K\K(m)$ for any $m\in M$ and
consequently the representation
$$\id:\A_\K\To\B(\M)$$ is irreducible.
\flushright\qed\end{proof}

In general case one can decompose the Hilbert space $\M=\bigoplus_{i=1}^N \M_i$, where $N\in\N$ or $N=\infty$, on the
invariant $\A_\K \M_i\subset\H_i$ orthogonal Hilbert subspaces. Superposing $\K:M\to\CP(\M)$ with the orthogonal projectors
$P_i:\H\to\H_i$ one obtains the family of coherent state maps $\K_i:=P_i\circ \K:M\to \CP(\M_i)$, $i=1,\ldots N$. One has
$\A_{\K_i}=P_i \A_\K P_i$ and the decomposition $\A_\K=\bigoplus_{i=1}^N A_{\K_i}$ is consistent with the decomposition
\be K_\alpha(m)=\sum_{i=1}^N (P_i\circ K_\alpha)(m),\quad m\in\Omega_\alpha\ee
of the coherent state map.

\begin{example}[\it Toeplitz Algebra]$\;$\label{ex:toeplitz}

Fix an orthonormal basis $\{|n\rangle \}_{n=1}^\infty$ in the Hilbert space $M$. The coherent states map $\K:\D\to\CP(\M)$ is
defined by \be \label{eq:Toep}\D\ni z\To\K(z):=\sum_{n=1}^\infty z^n|n\rangle  \ee where $\K(z)=[K(z)]$.

Quantum polarization $\overline\P_\K$ is generated in this case by the one-side shift operator
\be a|n\rangle =|n-1\rangle  \ee
which satisfies
\be aa^*=\I\ee
From this relation it follows that the algebra $\A_\K$ of physical observables generated by the coherent states map \eqref{eq:Toep} is Toeplitz $C^*$-algebra.
The existence  of normal ordering in $(\A_\K,\overline\P_\K)$ is guaranteed by the property that monomial
$$a^{*k}a^l\qquad k,l\in\N\cup\{0\} $$
are linearly dense in $\A_\K$.

Let us finally remark that the space $I(\M)$ is exactly the Hardy space $H^2(\D)$, see \cite{douglas,Rud}. According to the Theorem \ref{thm:irr} the auto-representation
of Toeplitz algebra is irreducible  as the unit disc $\D$ is connected and there exists the norm normal ordering in  $\A_\K$.
\end{example}

\begin{example}$\;$

Following \cite{Oqspec} one can generalize the construction presented in Example \ref{ex:toeplitz} taking
\be \label{R-coh}\D_\mathcal R\ni z\tto K_\mathcal R(z):=\sum_{n=1}^\infty\frac{z^n}{\sqrt{\mathcal R(q)\cdots \mathcal R(q^n)}}\cat n, \ee
where $0<q<1$ and $\mathcal R$ is a meromorphic function on $\C$ such that $\mathcal R(q^n)>0$ for $n\in\N\cup\{\infty\}$ and $\mathcal R(1)=0$.
For $z\in\D_\mathcal R:=\{z\in\C: \abs z< \sqrt{\mathcal R(0)}\}$ one has $K_\mathcal R(z)\in\M$ and the coherent state map
$\mathcal K_\mathcal R:\D_\mathcal R\to \CP(\M)$ is defined by $\K_\mathcal R(z)=\C K_\mathcal R(z)$. The annihilation $a$ and creation $a^*$
operators defined by \eqref{R-coh} satisfy the relations
\begin{align}
  a^*a&=\mathcal R(Q) \nonumber \\
  aa^*&=\mathcal R(qQ) \nonumber \\
  aQ&=qQa \nonumber \\
\label{R-alg}  Qa^*&=qa^*Q,
\end{align}
where the compact self-adjoint operator $Q$ is defined by $Q\cat n=q^n\cat n$. Hence one obtains the class of $C^*$-algebras
$\A_\mathcal R$ parametrized by the meromorphic functions $\mathcal R$, which includes the $q$-Heisenberg-Weyl algebra
of one degree of freedom and the quantum disc in sense of \cite{klimles} if
\be \mathcal R(x)=\frac{1-x}{1-q}\ee
and
\be \mathcal R(x)=r \frac{1-x}{1-\rho x},\ee
where $0<r,\rho\in\R$, respectively.

The algebras $\A_\mathcal R$ find application for the integration of quantum optical models, see \cite{HOT}.
For the rational $\mathcal R$ they also can be considered as the symmetry algebras in the theory of the basic hypergeometric series, see \cite{Oqspec}

In the paper \cite{O-L2} one investigates the case when $\mathcal R$ is invertible map. Then relations \eqref{R-alg} give
\be aa^*=\mathcal F(a^*a),\ee
where $\mathcal F:=\mathcal R\circ \mathcal L_q \circ \mathcal R^{-1}$ and $\mathcal L_q(x)=qx$. In this case the $C^*$-algebra
$\A_\mathcal R$ can be considered as the quantum algebra of the dynamical system defined by the function $\mathcal F$.
\end{example}

\begin{example}[\it q-Heisenberg-Weyl Algebra]$\;$\label{q-heis}

Let $M$ be the polydisc $\D_q\times\cdots\times\D_q$, where $\D_q\subset\C$ is the disc of radius $\frac{1}{\sqrt{1-q}}$, $0<q<1$. The orthonormal basis
in the Hilbert space $\M$ will be parameterized in the following way
$$\{|n_1\ldots n_N\rangle \}$$
where $n_1,\ldots,n_N\in\N\cup\{0\}$, and
$$\langle n_1\ldots n_N|k_1\ldots k_N\rangle =\delta_{n_1k_1}\ldots\delta_{n_Nk_N}$$
The coherent states map
$$\K:\D_q\times\cdots\times\D_q\To\CP(\M)$$
is defined by $\K(z_1,\ldots,z_N)=[K(z_1,\ldots,z_N)]$ where
\be \label{eq:Heis}K(z_1,\ldots,z_N):=\sum_{k_1,\ldots,k_N=0}^\infty \frac{z_1^{k_1}\cdots z_N^{k_N}}{\sqrt{[k_1]!_q\cdots[k_N]!_q}}|k_1\ldots k_N\rangle  \ee
The standard notation
$$[n]:=1+\cdots+q^{n-1}$$
$$[n]!_q:=[1]\cdots[n]$$
was used in \eqref{eq:Heis}.

The quantum polarization $\overline\P_\K$ is the algebra generated by the operators $a_1,\ldots,a_N$ defined by
\be a_i K(z_1,\ldots,z_N)=z_iK(z_1,\ldots,z_N) \ee
It is easy to show that $\norm{a_i}=\frac{1}{\sqrt{1-q}}$. Hence $\overline\P_\K$ is commutative and algebra $\A_\K$ of all quantum observables is generated
by $\I,a_1,\ldots,a_N,a_1^*,\ldots,a_N^*$ satisfying the relations
$$[a_i,a_j]=[a_i^*,a_j^*]=0$$
\be \label{eq:comHeis}a_ia_j^*-qa_j^*a_i=\delta_{ij}\I. \ee
The $C^*$-algebra $\A_\K$ is then the q-deformation of Heisenberg-Weyl algebra, see \cite{jorgensen}. The structural relations \eqref{eq:comHeis} imply that
$a_i^{*k}a_j^l$, where $i,j=1,\ldots,N$ and $k,l\in\N\cup\{0\}$ do form linearly dense subset in $\A_\K$. Consequently $\A_\K$ admits the norm normal ordering. Since
the polydisc is connected the auto-representation of $\A_\K$ is irreducible.

In the limit $q\to1$ $\A_\K$ becomes the standard Heisenberg-Weyl algebra for which the creation and annihilation operators
are unbounded.
\end{example}

\begin{example}[\it Quantum Complex Minkowski Space]$\;$

    For detailed investigation of quantum complex Minkowski space see \cite{OJ}. In this case the classical phase space is the symmetric domain
\be \mathbb D:=\{Z\in Mat_{2\times 2}(\C) : E-Z^\dag Z > 0 \},\ee
where $E=\left(%
\begin{array}{cc}
  1 & 0 \\
0 & 1 \\\end{array}%
\right)$. The coherent state map $\K_\lambda=[K_\lambda]:\mathbb D\to\CP(\M)$, $\N\ni \lambda>3$, is given by
\be\label{coh}
K_\lambda: Z \to |Z;\lambda\rangle:=\sum_{j,m,j_{1},j_{2}}\Delta^{jm}_{j_{1}j_{2}}(Z)\left|\begin{array}{@{}cc@{}}
  j & m \\
  j_{1}& j_{2} \\
\end{array}\right\rangle,\ee
where
\be \Delta^{jm}_{j_{1}j_{2}}(Z):=(N^{\lambda}_{jm})^{-1}(\det Z)^{m}\sqrt{\frac{(j+j_{1})!(j-j_{1})!}{(j+j_{2})!(j-j_{2})!}}\times\ee
$$\times\!\!\!\sum_{\substack{S\geq \max\{0,j_{1}+j_{2}\}\\ S\leq \min\{j+j_{1},j+j_{2}\}}}
\!\!\!\!\left(%
\begin{array}{@{}c@{}}
  j+j_{2} \\
  S \\
\end{array}%
\right)\left(%
\begin{array}{@{}c@{}}
  j-j_{2}\\
  S-j_{1}-j_{2} \\
\end{array}%
\right)z_{11}^{S}z_{12}^{j+j_{1}-S}z_{21}^{j+j_{2}-S}z_{22}^{S-j_{1}-j_{2}}$$
and
\be
N^{\lambda}_{jm}:=(\lambda-1)(\lambda-2)^{2}(\lambda-3)\frac{\Gamma(\lambda-2)\Gamma(\lambda-3)m!(m+2j+1)!}{(2j+1)!\Gamma(m+\lambda-1)\Gamma(m+2j+\lambda)}.\ee

By
\be\label{baza}\left\{\left|\begin{array}{@{}cc@{}}
  j & m \\
  j_{1} & j_{2} \\
\end{array}\right\rangle\right\},\ee
where $m,2j\in\mathbb{N}\cup\{0\}$ and $-j\leq j_{1},j_{2}\leq j$, we denote an orthonormal basis
in $\mathcal M$, i.e.
\be \left\langle \begin{array}{@{}cc@{}}
  j & m \\
  j_{1} & j_{2} \\
\end{array}\right.\left|\begin{array}{@{}cc@{}}
  j' & m' \\
  j_{1}' & j_{2}' \\
\end{array}\right\rangle
=\delta_{jj'}\delta_{mm'}\delta_{j_1j_1'}\delta_{j_2j_2'}.\ee

The quantum polarization $\overline\P_{\K_\lambda}$ is generated by the following four annihilation operators
\begin{eqnarray} a_{11}\left|\begin{array}{@{}cc@{}}
  j & m \\
  j_{1}& j_{2} \\
\end{array}\right\rangle
&=\sqrt{\frac{(j-j_{1}+1)(j-j_{2}+1)m}{(2j+1)(2j+2)(m+\lambda-2)}}\left|\begin{array}{@{}cc@{}}
  j+\frac{1}{2} & m-1 \\
  j_{1}-\frac{1}{2}& j_{2}-\frac{1}{2} \\
\end{array}\right\rangle \nonumber \\ &+\sqrt{\frac{(j+j_{1})(j+j_{2})(m+2j+1)}{(m+2j+\lambda-1)2j(2j+1)}}\left|\begin{array}{@{}cc@{}}
  j-\frac{1}{2} & m \\
  j_{1}-\frac{1}{2}& j_{2}-\frac{1}{2} \\
\end{array}\right\rangle\end{eqnarray}

\begin{eqnarray} a_{12}\left|\begin{array}{@{}cc@{}}
  j & m \\
  j_{1}& j_{2} \\
\end{array}\right\rangle
&=-\sqrt{\frac{(j-j_{1}+1)(j+j_{2}+1)m}{(2j+1)(2j+2)(m+\lambda-2)}}\left|\begin{array}{@{}cc@{}}
  j+\frac{1}{2} & m-1 \\
  j_{1}-\frac{1}{2}& j_{2}+\frac{1}{2} \\
\end{array}\right\rangle \nonumber \\
&+\sqrt{\frac{(j+j_{1})(j-j_{2})(m+2j+1)}{(m+2j+\lambda-1)2j(2j+1)}}\left|\begin{array}{@{}cc@{}}
  j-\frac{1}{2} & m \\
  j_{1}-\frac{1}{2}& j_{2}+\frac{1}{2} \\
\end{array}\right\rangle\end{eqnarray}

\begin{eqnarray} a_{21}\left|\begin{array}{@{}cc@{}}
  j & m \\
  j_{1}& j_{2} \\
\end{array}\right\rangle
&=-\sqrt{\frac{(j+j_{1}+1)(j-j_{2}+1)m}{(2j+1)(2j+2)(m+\lambda-2)}}\left|\begin{array}{@{}cc@{}}
  j+\frac{1}{2} & m-1 \\
  j_{1}+\frac{1}{2}& j_{2}-\frac{1}{2} \\
\end{array}\right\rangle \nonumber \\ &+\sqrt{\frac{(j-j_{1})(j+j_{2})(m+2j+1)}{(m+2j+\lambda-1)2j(2j+1)}}\left|\begin{array}{@{}cc@{}}
  j-\frac{1}{2} & m \\
  j_{1}+\frac{1}{2}& j_{2}-\frac{1}{2} \\
\end{array}\right\rangle\end{eqnarray}

\begin{eqnarray} a_{22}\left|\begin{array}{@{}cc@{}}
  j & m \\
  j_{1}& j_{2} \\
\end{array}\right\rangle
&=\sqrt{\frac{(j+j_{1}+1)(j+j_{2}+1)m}{(2j+1)(2j+2)(m+\lambda-2)}}\left|\begin{array}{@{}cc@{}}
  j+\frac{1}{2} & m-1 \\
  j_{1}+\frac{1}{2}& j_{2}+\frac{1}{2} \\
\end{array}\right\rangle \nonumber \\
&+\sqrt{\frac{(j-j_{1})(j-j_{2})(m+2j+1)}{(m+2j+\lambda-1)2j(2j+1)}}\left|\begin{array}{@{}cc@{}}
  j-\frac{1}{2} & m \\
  j_{1}+\frac{1}{2}& j_{2}+\frac{1}{2} \\
\end{array}\right\rangle. \end{eqnarray}
In the expressions above we put by definition $\left|\begin{array}{@{}cc@{}}
  j & m \\
  j_1& j_2 \\
\end{array}\right\rangle:=0$ if the indices do not satisfy the condition $m,2j\in\mathbb{N}\cup\{0\}$ and $-j\leq j_{1},j_{2}\leq j$.

The quantum symmetric domain $\A_{K_\lambda}$ is an operator $C^*$-algebra containing the ideal $L^0(\M)$ of compact
operators in such way that $L^0(\M)\cap \overline\P_{\K_\lambda}=\{0\}$ and $L^0(\M)\subsetneq Comm \A_{\K_\lambda}$.
One has the isomorphism of $\A_{\K_\lambda}/Comm \A_{\K_\lambda}$ with the algebra of continuous functions $C(U(2))$
on the \v Silov
boundary $U(2):=\{Z\in Mat_{2\times2}(\C) \;:\; ZZ^*=E\}$ of $\D$.
After application of the Caley transform one shows that $U(2)$ is conformal compactification of Minkowski space and thus
$\A_{\K_\lambda}$ has the interpretation of quantum Minkowski space.

\end{example}

\section{Coherent states and multiplicative functionals}\label{4}
In this section we will study the properties of the space
$\c(\A,\overline\P)$ of coherent states on the abstract polarized
$C^*$-algebra $(\A,\overline\P)$ with admissible norm normal ordering.
The coherent state $\omega\in\c(\A,\overline\P)$, after
restriction to $\overline\P$, becomes the multiplicative linear
functional
$$\omega_{|\overline\P}:\overline\P\To\C$$
Consequently one has the map
\be \rho: \c(\A,\overline\P) \To M(\overline\P) \ee
defined by the restriction $\rho(\omega):=\omega_{|\overline\P}$,
which maps the space of coherent states
$\c(\A,\overline\P)$ into the space of the multiplicative linear
functionals on the commutative Banach algebra $\overline\P$. From
the inequality
\be \norm{\rho(\omega)}\leq\norm\omega \ee
and from the existence of the normal ordering on
$(\A,\overline\P)$ it follows that $\rho$ is an injective
contraction. This allows us to identify $\c(\A,\overline\P)$ with
the subspace $\rho\left(\c(\A,\overline\P)\right)$ of the space
$M(\overline\P)$ of all multiplicative functionals on
$\overline\P$.

The theorem presented below characterizes multiplicative
functionals from the image $\rho\left(\c(\A,\overline\P)\right)$.
\begin{theorem}\label{thm:10}
The multiplicative functional $\mu\in M(\overline\P)$ can be
prolongated to the coherent state on $(\A,\overline\P)$ iff for
any $a_1,\ldots,a_J,b_1,\ldots,b_S\in\overline\P$ one has
\be \label{eq:51}\left(\sum_{j=1}^J a_j^*a_j\leq\sum_{s=1}^S b_s^*b_s\right)\Rightarrow
\left(\sum_{j=1}^J \abs{\mu(a_j)}^2\leq\sum_{s=1}^S
\abs{\mu(b_j)}^2\right). \ee
\end{theorem}
\begin{proof}
The elements
\be \label{eq:52}x=\sum_{j=1}^Jd_j^*c_j \ee
where $c_i,d_j\in\overline\P$, form the *-invariant complex vector
subspace $V$ in it. Let multiplicative functional $\mu$ satisfy
the condition \eqref{eq:51}. We will define the linear functional
$\omega:V\to\C$ by the formula
\be \label{eq:53}\omega\left(\sum_{j=1}^J
d_j^*c_j\right):=\sum_{j=1}^J\overline{\mu(d_j)}\mu(c_j).\ee
In order to show that this definition of $\omega$  is correct we
notice that
\be d_j^*c_j=\frac{1}{4}\sum_{k=0}^3i^k(c_j+id_j)^*(c_j+id_j). \ee
Any $x=x^*\in V$ can be thus expressed in the form
$$x=\sum_{j=1}^Ja_j^*a_j-\sum_{s=1}^Sb_s^*b_s. $$
If one takes another decomposition
$$x=\sum_{j=1}^J{a'}_j^*{a'}_j-\sum_{s=1}^S{b'}_s^*b'_s. $$
then due to the equality
$$\sum_{j=1}^Ja_j^*a_j+\sum_{s=1}^S{b'}_s^*b'_s=\sum_{j=1}^J{a'}_j^*a'_j+\sum_{s=1}^Sb_s^*b_s$$
and from \eqref{eq:51} it follows that definition \eqref{eq:53}
does not depend on the presentation \eqref{eq:52} of the Hermitean
element $x^*=x\in V$. It is clear that
$\omega(x^*)=\overline{\omega(x)}$ for $x\in V$. So, $\omega$ is
well defined on $V$. The condition \eqref{eq:51} guarantees its
positivity. Hence, according to the Lemma 2.10.1 in \cite{dixmier}, $\omega$
can be prolongated from $V$ to a possitive functional on $\A$.
Directly from definition \eqref{eq:53} it follows that $\omega$
does satisfy the condition \eqref{eq:coh} for $x\in V$ and
$a\in\overline\P$. Since $\omega$ is continuous and $V$ is dense
in $\A$ the condition \eqref{eq:coh} is valid for any $x\in\A$. We
have proven that $\mu$ can be uniquely prolongaded to the coherent
state $\omega\in\c(\A,\overline\P)$.

If $\omega\in\c(\A,\overline\P)$ then due to its positivity and the condition \eqref{eq:coh} one gets the property
\eqref{eq:51}. \flushright\qed\end{proof}

From the above one may draw the following
\begin{corollary} \label{cor:pure}The image $\rho\left(\c(\A,\overline\P)\right)$ is the
compact subset (in the sense of weak topology) in the space
$M(\overline\P)$ of multiplicative linear functionals on the
commutative Banach algebra $\overline\P$. \end{corollary}
\begin{proof}If $\mu_n\to\mu\in M(\overline\P)$ in weak topology,
meaning that $\mu_n(a)\to\mu(a)$ for any $a\in M(\overline\P)$, then $\mu$ satisfies the condition \eqref{eq:53}
provided $\mu_n$ are subject to the same condition. According to Theorem \ref{thm:10}
$\mu\in\rho\left(\c(\A,\overline\P)\right)$ if $\mu_n\in\rho\left(\c(\A,\overline\P)\right)$. Hence
$\rho\left(\c(\A,\overline\P)\right)$ is closed subset of the compact set $M(\overline\P)$ and must be compact too.
\flushright\qed\end{proof}

We shall fix some notation related with GNS construction. Let $\omega$ be the state on $C^*$-algebra $\A$. By
$N_\omega$ we will denote the left sided ideal in $\A$ containing all elements $x\in\A$ such that $\omega(x^*x)=0$. By
\be \label{eq:55} \norm{[x]}_\omega^2=\langle [x]|[x]\rangle _\omega:=\omega(x^*x) \ee we shall denote the scalar product of
equivalence class \be \label{eq:56} [x]:=x+N_\omega\in \A/N_\omega=\M_\omega \ee with itself.

The Hilbert space being the closure of $\A/N_\omega$ in the norm
$\norm\cdot_\omega$ will be denoted by $\H_\omega$. By $v_\omega$
we shall denote the element $[\I]\in\H_\omega$. From the
Theorem \ref{thm:coh} it follows
\begin{proposition} Let $\omega$ be the coherent state on $(\A,\overline\P)$. Then:

\begin{enumerate}[i)]
\item \be\ker \omega=N_\omega+N_\omega^*, \ee
where $\ker \omega = \{x\in\A : \omega(x)=0\}$.
\item The GNS representation
$$ \pi_\omega:\A\To \End\H_\omega $$
of $C^*$-algebra $\A$ is irreducible.
\end{enumerate}
\end{proposition}
We shall formulate and prove now the theorem which gives
equivalent criteria for the state to be a coherent one.
\begin{theorem} \label{thm:13}
Let $\omega$ be the state on the polarized algebra
$(\A,\overline\P)$. The following conditions:
\begin{enumerate}[1)]
\item $\omega\in\c(\A,\overline\P)$,
\item $N_\omega\cap\overline\P=\ker\omega\cap\overline\P$,
\item $\pi_\omega(a)v_\omega=\omega(a)v_\omega$ for
$a\in\overline\P$,
\item $N_\omega\cap\overline\P$ is the maximal ideal in
$\overline\P$.
\end{enumerate}
are equivalent.
\end{theorem} \begin{proof}

$1) \Rightarrow 2)$

The element $x\in N_\omega$ iff $\pi_\omega(x)v_\omega=0$. Hence
$$\omega(a)=\langle v_\omega|\pi_\omega(a)v_\omega\rangle =0$$
for $a\in N_\omega\cap \overline\P$. Consequently
$N_\omega\cap\overline\P\subset\ker\omega\cap\overline\P$.
Conversely, if $a\in\ker\omega\cap\overline\P$ then
$$0=\abs{\omega(a)}^2=\omega(a^*a),$$
meaning that $\ker\omega\cap \overline\P\subset
N_\omega\cap\overline\P$.

$2) \Rightarrow 3)$

If $2)$ is satisfied then $a-\omega(a)\I\in
N_\omega\cap\overline\P$. This in turn implies that
$$\pi_\omega\left(a-\omega(a)\I\right)v_\omega=0 $$
for any $a\in\overline\P$ and consequently the condition $3)$.

$3) \Rightarrow 1)$

In GNS representation one has
$$\omega(xa)=\langle v_\omega|\pi_\omega(xa)v_\omega\rangle =\langle v_\omega|\pi_\omega(x)\omega(a)v_\omega\rangle =\omega(x)\omega(a)$$
for $x\in\A$ and $a\in\overline\P$. The state is thus coherent.

We have shown the equivalence of 1), 2), 3).

$1)\Rightarrow 4)$

If $\omega\neq 0$ is the coherent state then $\rho(\omega)$ is the
multiplicative functional on $\overline\P$. The intersection
$\ker\omega\cap\overline\P$ is thus the maximal ideal in
$\overline\P$. Since $1)\Leftrightarrow 2)$ the set
$N_\omega\cap\overline\P$ is the maximal ideal in $\overline\P$
too.

$4)\Rightarrow 3)$

Assume that $N_\omega\cap \overline\P$ is the maximal ideal in
$\overline\P$. One then has
$$\overline\P=\left(N_\omega\cap\overline\P\right)\oplus\C\I$$
and consequently for any $a\in\overline\P$ there exists $\alpha\in\C$ such that $a-\alpha\I\in N_\omega\cap
\overline\P$. This implies
\be \label{eq:59} \pi_\omega(a-\alpha\I)v_\omega=0. \ee
The equality
\be \omega(a)=\langle v_\omega|\pi_\omega(a)v_\omega\rangle =\alpha\ee
together with \eqref{eq:59} gives 3). This shows
$4)\Rightarrow3)$. This way we have shown the equivalence of all conditions of Theorem \ref{thm:13}.
\flushright\qed\end{proof}

From the equivalence of the conditions 1) and 3) of the theorem above one immediately obtains the following corollary.
\begin{corollary} \label{cor:14}
For any coherent state $\omega$ on $(\A,\overline\P)$ the vector
space
\be \pi_\omega(\P)v_\omega=\left\{\pi_\omega(a)v_\omega :
a\in\P\right\} \ee
is dense in the Hilbert space $\H_\omega$ of GNS representation.
\end{corollary}

The property  described above indicates that the formalism
developed in this section is proper generalization of the Fock
representation in quantum mechanics and in quantum field theory. This
also justifies the interpretation of the commutative algebra $\P$
as the algebra of creation operators.

Let $(\pi_\omega,\H_\omega,v_\omega)$ and $(\pi_\nu,\H_\nu,v_\nu)$
be the pair of irreducible GNS representations corresponding to
the coherent states $\omega$ and $\nu$. The unitary equivalence of
representations $\pi_\omega\sim\pi_\nu$ defines the equivalence
relation of the coherent states $\omega\sim\nu$. By
$[\omega]\subset\c(\A,\overline\P)$ we shall denote the
equivalence class of coherent states equivalent to $\omega$ in the
sense above.
\begin{lemma}\label{lem:15}
Let $(\pi_\omega,\H_\omega,v_\omega)$ and $(\pi_\nu,\H_\nu,v_\nu)$
be GNS representations generated by the coherent states $\omega$ and
$\nu$ respectively. One then has
\begin{enumerate}[i)]
\item $\big(\omega\sim\nu\big)\Leftrightarrow\big(\exists
v\in\H_\omega \textrm{ such that }\nu(x)=\langle v|\pi_\omega(x)v\rangle  \textrm{ for
any } x\in\A\big)\Leftrightarrow\big(\exists
w\in\H_\nu \textrm{ such that }\omega(x)=\langle w|\pi_\nu(x)w\rangle  \textrm{ for
any } x\in\A\big)$;
\item The vector state
\be \label{eq:62}\nu(x)=\langle v|\pi_\omega(x)v\rangle , \ee
where $v\in\H_\nu$ is coherent state iff
\be \label{eq:63}\pi_\omega(a)v=\nu(a)v \ee
for any $a\in\overline\P$.
\end{enumerate}
\end{lemma}
\begin{proof}
$\quad$
\begin{enumerate}[i)]
\item Obvious
\item From \eqref{eq:62} and the property \eqref{eq:coh}
one obtains
\be \label{eq:64} \langle \pi_\omega(x^*)v|(\pi_\omega(a)-\nu(a)\I)v>=0. \ee
The set of vectors $\pi_\omega(x^*)v$, $x\in\A$, is dense in $\H_\omega$ as the representation $\pi_\omega$ is
irreducible. The equality \eqref{eq:64} implies then \eqref{eq:63}. Taking the scalar product of $\pi_\omega(x^*)v$ with \eqref{eq:63} one shows the property \eqref{eq:coh} for $\nu$.
\end{enumerate}
\flushright\qed\end{proof}

\begin{lemma} \label{lem:2}
$\quad$
\begin{enumerate}[i)]
\item Let $\omega\in\c(\A,\overline\P)$. Then
\be \label{eq:65} \ker\pi_\omega\cap\P=N_\omega\cap \P. \ee
\item If $\omega,\nu\in\c(\A,\overline\P)$ are equivalent then one has the equality
\be \label{eq:66} N_\omega\cap\P=N_\nu\cap\P \ee
of the corresponding ideals.
\end{enumerate}
\end{lemma}
\begin{proof}
$\quad$
\begin{enumerate}[i)]
\item According to GNS construction the left sided ideals $N_\omega$ consists of those $x\in\A$ which
annihilate the vector $v_\omega\in\H_\omega : \pi_\omega(x)v_\omega=0$. This implies
$$\ker \pi_\omega\cap \P\subset N_\omega\cap \P.$$
Since $N_\omega\cap\P$ is an ideal in $\P$ one has
$$\pi_\omega(a)\pi_\omega(b^*)v_\omega=0$$
for any $a\in N_\omega\cap\P$ and any $b\in\overline\P$. The vectors $\pi_\omega(b^*)$, $b\in\overline\P$,
according to the Corollary \ref{cor:14} do constitute the dense subspace in $\H_\omega$. One thus has
$$N_\omega\cap\P\subset\ker\pi_\omega\cap\P.$$
\item The equivalence $\omega\sim\nu$ of the coherent states $\omega$ and $\nu$ implies the equality of the
kernels
$$\ker\pi_\omega=\ker\pi_\nu$$
of the corresponding representations. From this together with i) one obtains \eqref{eq:66}.
\end{enumerate}
\flushright\qed\end{proof}

The proposition below gives necessary and sufficient condition for the coherent states to be equivalent.
\begin{proposition} \label{prop:17}
Let $\omega$ and $\nu$ be the coherent states and $(\pi_\omega,\H_\omega,v_\omega)$ $(\pi_\nu,\H_\nu,v_\nu)$ be
their respectively their GNS representations. Then the following conditions:
\ben[(i)]
\item $\omega\sim\nu$
\item $\dim (\pi_\omega(\ker\nu\cap \P)v_\omega)^\perp = 1$
\item $\dim (\pi_\nu(\ker\omega\cap \P)v_\nu)^\perp = 1$
\een
are equivalent.
\end{proposition}
\begin{proof}
Assume that
$$\dim\big(\pi_\omega(\ker\nu\cap\P)v_\omega\big)^\perp=1.$$
Then there exists a vector $v\in\big(\pi_\omega(\ker\nu\cap\P)v_\omega\big)^\perp$ such that
$\norm{v}=1$ and
\be\label{eq:68}
0=\langle v|\pi_\omega(\ker\nu\cap\P)\pi_\omega(\P)v_\omega\rangle =\langle \pi_\omega(\ker\nu\cap\overline\P)v|\pi_\omega(\P)
v_\omega\rangle  .\ee
The equality \eqref{eq:68} is true due to the fact that $\ker\nu\cap\P$ is an ideal in $\P$. From
\eqref{eq:68} and because $\pi_\omega(\P)v_\omega$ is dense in $\H_\omega$ one has
\be \label{eq:69} \pi_\omega(a-\nu(a)\I))v=0 \ee
for any $a\in\overline\P$. One can thus represent the coherent state $\nu$ by
\be \label{eq:70} \nu(x)=\langle v|\pi_\omega(x)v\rangle  \ee
for $x\in\A$. According to Lemma \ref{lem:15} and \eqref{eq:69}, \eqref{eq:70} the coherent state $\nu$ is
equivalent to $\omega$.

Conversely, if one assumes $\nu\sim\omega$, the again due to ii) of Lemma \ref{lem:15} there exists
$v\in\H_\omega$, $\norm{v}=1$ such that
$$\nu(x)=\langle v|\pi_\omega(x)v\rangle $$
for $x\in\A$, and
\be \label{eq:71} \pi_\omega(a)v=0 \ee
for $a\in \ker\nu\cap\overline\P$. Thus one has
$$v\in\big(\pi_\omega(\ker\nu\cap\P)v_\omega\big)^\perp,$$
showing that $\big(\pi_\omega(\ker\nu\cap\P)v_\omega\big)^\perp\neq\{0\}$. As the relation $\omega\sim\nu$ is symmetric
the conditions $\big(\pi_\omega(\ker\nu\cap\P)v_\omega\big)^\perp\neq\{0\}$ and
$\big(\pi_\nu(\ker\omega\cap\P)v_\nu\big)^\perp\neq\{0\}$ are equivalent. Since $\pi_\omega(\P)v_\omega$ is linearly
dense in $\H_\omega$ and codimension of $\ker \nu\cap\P$ in $\P$ is equal one we have $\dim (\pi_\omega(\ker\nu\cap
\P)v_\omega)^\perp = 1$. This ends the proof. \flushright\qed\end{proof}

The considerations above shows that one can describe the coherent states on $(\A,\overline\P)$ in terms of
multiplicative functionals on the polarization $\overline\P$, or equivalently, on the antipolarization $\P$. Let us recall
that by
definition $\P$ consists of the elements *-conjugated to those of $\overline\P$.

Now, we will show the similar possibility on the level of GNS representations $(\pi_\omega,\M_\omega,v_\omega)$ generated by the coherent states
$\omega\in\C(\A,\overline\P)$. In order to do this let us consider the quotient vector space $\P/\P\cap N_\omega$. The coherent state $\omega$
defines the scalar product
\be \label{eq:88} \langle [a]|[b]\rangle _\omega:=\omega(a^*b) \ee
on the vectors $[a],[b]\in\P/\P\cap N_\omega$. The completion of $\P/\P\cap N_\omega$ in the norm $\norm\cdot_\omega$ given by the scalar
product \eqref{eq:88} is the Hilbert space $H_\omega$. This space is evidently defined in terms of antipolarization $\P$ and multiplicative
functionals $\omega_{|\P}$ given by restriction of $\omega$ to $\P$.
Since $\P\cap N_\omega$ is the ideal of commutative algebra $\P$ one has the representation $\gamma_\omega:\P\to\End(\P/\P\cap N_\omega)$
of $\P$ defined by
\be \label{eq:89} \gamma_\omega(a)[b]=[ab] \ee
It follows from $\norm{\gamma_\omega(a)}_\omega\leq\norm a$ that $\gamma_\omega$ extends to the representation of commutative Banach algebra $\P$
in the Hilbert space $H_\omega$. Taking care of the normal ordering in $(\A,\overline\P)$ we can extend $\gamma_\omega$ onto whole $C^*$-algebra
$\A$ by putting
\be \label{eq:89a} \gamma_\omega(x):=\sum_{k=1}^N\gamma_\omega(b_k)\gamma_\omega(a_k)^* , \ee
where $x=\sum_{k=1}^Nb_ka_k^*$ and $b_k,a_k\in\P$.

We will show that in the case of $\A$ being Toeplitz algebra the Hilbert space $H_\omega$ is naturally isomorphic to the Hardy space $H^2(\D)$.
The representation $\gamma_\omega$ is the the natural representation of Toeplitz algebra generated by $H^\infty(\D)=\P$.
Being motivated by this important example we shall call the representation
\be \label{eq:90} \gamma_\omega:\A\To\B(H_\omega) \ee
the {\bf Hardy type representation} of the polarized $C^*$-algebra $(\A,\overline\P)$.

The map
\be \label{eq:91}U_\omega([b]):=\pi_\omega(b)v_\omega, \ee
where $b\in\P$, defines an isometry of the unitary space $\P/\P\cap N_\omega$ into the Hilbert space $\H_\omega$. From Corollary \ref{cor:14}
it follows that one can extend $U_\omega$ to the unitary isomorphism $U_\omega:H_\omega\to\H_\omega$ of the Hilbert spaces since
$$U_\omega\gamma_\omega(a)([b])=U_\omega([ab])=\pi_\omega(ab)v_\omega=\pi_\omega(a)(\pi_\omega(b)v_\omega)=$$
\be \label{eq:92} =\pi_\omega(a)\circ U_\omega([b]) \ee
for any $b\in\P$. Thus we have
$$U_\omega\circ\gamma_\omega(a)=\pi_\omega(a)\circ U_\omega $$
and
\be \label{eq:94} U_\omega\circ\gamma_\omega^*(a)=\pi_\omega(a^*)\circ U_\omega. \ee
From \eqref{eq:89a} and \eqref{eq:94} we conclude now that $U_\omega$ intertwine Hardy representation with GNS representation. In order to summarize we
formulate
\begin{proposition} \label{prop:19}
The Hardy representation $(\gamma_\omega,H_\omega)$ is equivalent to GNS representation $(\pi_\omega,\H_\omega,v_\omega)$.
\end{proposition}
In order to stress this property the abbreviated notation HGNS will be used for the representation of Hardy type.

The construction presented above is thus the natural generalization of the construction of Fock representation for the
Heisenberg algebra constructed with creation operators.

\section{Classical and quantum K\"ahler structures}\label{5}
We are now in a position to reconstruct the coherent states map
\be \label{eq:74} \K_\omega:M_\omega\to\CP(\H_\omega), \ee
out of the coherent  state $\omega$. The set $M_\omega:=[\omega]\subset M(\overline\P)$ will be equipped with weak topology. By
$\H_\omega$ we will as usually denote the Hilbert space of the GNS representation
$(\pi_\omega,\H_\omega,v_\omega)$.

According to the Proposition \ref{prop:17} if $v\in M_\omega$ then the dimension of the vector subspace
$$\big(\pi_\omega(\ker\nu\cap\P)v_\omega\big)^\perp\subset\H_\omega$$
is equal to one. In addition the vector
$$v\in\big(\pi_\omega(\ker\nu\cap\P)v_\omega\big)^\perp$$
has the property \eqref{eq:63} of Lemma \ref{lem:15}. Hence, the map $\K_\omega$ given by
\be \label{K_om}M_\omega\ni v\To\K_\omega(v):=\big(\pi_\omega(\ker\nu\cap\P)\big)^\perp\in\CP(\H_\omega)\ee
satisfies the property \eqref{eq:anih} for the elements $\pi_\omega(a)$ of the commutative operator
algebra $\pi_\omega(\overline\P)$. The common eigenvalues function for the algebra $\pi_\omega(\overline\P)$
is given by
\be \label{eq:76} \mn a(\nu):=\nu(a)=\langle v_\nu|\pi_\omega(a)v_\nu\rangle. \ee
It equals to the restriction $\hat{a}_{|M_\omega}$ of the Gelfand transform image $\hat a:M(\overline\P)\to\C$ of the element
$a\in\overline\P$.

The above justifies one to call the map $\K_\omega$ the {\bf coherent states map related} to
$\omega\in\c(\A,\overline\P)$. If one takes the coherent state $\omega^u$ equivalent to $\omega$ then there
exists unitary element $u\in\A$ such that
\be \label{eq:77} \omega^u(x)=\omega(u^*xu) \ee
for any $x\in\A$ (see \cite{dixmier,murphy}). Then there also exists the unitary isometry $U:\H_\omega\to\H_{\omega^u}$ which is
defined by
\be \label{eq:78} U(\pi_\omega(x)v_\omega):=\pi_{\omega^u}(x)v_{\omega^u} \ee
such that
\be \label{eq:79} \K_{\omega^u}=[U]\circ\K_\omega,\ee
where $[U]:\CP(\H_\omega)\to\CP(\H_{\omega^u})$ is the projectivisation of $U$.

In the proof of the theorem below we shall identify all equivalent Hilbert spaces $\H_{\omega^u}$ with
$\H_\omega$ and the map $U$ with $\pi_\omega(u)$. The notation of \eqref{eq:55} and \eqref{eq:56} will be in
use.
\begin{theorem}
Assume that the GNS representation $(\pi_\omega,\H_\omega,v_\omega)$ does satisfy the condition
\be \label{eq:80} \pi_\omega(\A)\cap L^0(\H_\omega)\neq\{0\}, \ee
where $L^0(\H_\omega)$ is the ideal of all compact operators in $\H_\omega$. Then the coherent states map
$\K_\omega:M_\omega\to\CP(\H_\omega)$ is continuous.
\end{theorem}
\begin{proof}
Let us take the sequence of the coherent states $\omega'_n\in M_\omega$, $n\in\N$, converging to $\omega'\in
M_\omega$. Since $\omega'$ and $\omega'_n$ are pure states there are corresponding unitary elements $u$ and
$u_n$ from $\A$ such that
$$\omega'(x)=\omega(u^*xu) $$
\be \label{eq:81} \omega'_n(x)=\omega(u_n^*xu_n) \ee
for any $x\in\A$. One then has
$$\K_\omega(\omega')=\big[[u]\big] $$
and
\be \label{eq:82} \K_\omega(\omega'_n)=\big[[u_n]\big]\ee
In order to show the convergence
\be \label{eq:83} \big[[u_n]\big]\to\big[[u]\big] \ee
for $\omega'_n\to\omega'$, we use the equality
$$\norm{[u_n]-[u]}^2=\langle [u_n]_\omega-[u]|[u_n]-[u]\rangle =$$
\be \label{eq:84}=\omega(u_n^*u_n)+\omega(u^*u)-\omega(u^*u_n)-\omega(u_n^*u)= 2-\omega(u^*u_n)-\omega(u^*_nu)
.\ee Let us take now the projector $\omega(u^*\cdot)[u]\in L^0(\H_\omega)$. From \eqref{eq:80} it follows
that $L^0(\H_\omega)\subset\pi_\omega(\A)$ (see \cite{murphy}). So, there is $x_0\in\A$ such that
\be  \pi_\omega(x_0)=\omega(u^*\cdot)[u].\ee
This gives in turn
$$\omega'_n(x_0)=\omega(u_n^*x_0u_n)=\langle v_\omega|\pi_\omega(u_n^*x_0u_n)v_\omega\rangle =\langle [u_n]|\omega(u^*\cdot)[u][u_n]\rangle =$$
\be \label{eq:87} =\omega(u^*u_n)\langle [u_n]|[u]\rangle =\abs{\omega(u^*u_n)}^2 \ee
and
$$\omega'(x_0)=\abs{\omega(u^*u)}^2=1 $$
The unitary elements $u$ and $u_n$ could be chosen up to the phase factors $e^{i\phi}$ and $e^{i\phi_n}$, where  $\phi$,$\phi_n\in\R$. So
one can assume $\omega(u^*u_n)\geq 0$. From this and from $\omega'_n(x_0)\to\omega'(x_0)$ we conclude that
$$\omega(u^*u_n)\to 1.$$
Hence
$$\norm{[u_n]-[u]}^2\to 0,$$
what implies the convergence of \eqref{eq:83}. \flushright\qed\end{proof}

If one assumes that it is a postlimital $C^*$-algebra (see \cite{murphy}) then the condition \eqref{eq:81} is satisfied for any coherent state
$\omega\in\c(\A,\overline\P)$. Consequently for postlimital $C^*$-algebras the coherent states maps $\K_\omega$ are always continuous.

Now let us consider the commutative Banach algebra homomorphism
\be \Phi_\omega:\P\ni a^*\tto \overline{\mn a}\in C(M_\omega),\ee
where $\C(M_\omega)$ is the algebra of continuous functions on $M_\omega$ and $\mn a$ is defined by \eqref{eq:76}. According
to Lemma \ref{lem:2} one has $\ker \Phi_\omega=\ker \pi_\omega \cap \P=N_\omega\cap\P$. Hence
$\ker \Phi_\omega$ depends on $[\omega]$ only and $\Phi_\omega$ defines the isomorphism  of quotient algebra $\P/N_\omega\cap\P$ with function algebra
$\im \Phi_\omega\subset C(M_\omega)$.

The coherent state map $\K_\omega:M_\omega\to \CP(\H_\omega)$ given by \eqref{K_om} defines
line bundle $\K_\omega^*\E=:\L\to M_\omega$ and the anti-linear monomorphism $I:\H_\omega\to\Gamma(M,\overline{\L^*})$
of complex vector spaces in the way describe in Section \ref{2}. For $\lambda\in\im\Phi_\omega$ and $\psi\in I(\H_\omega)$
one has $\lambda\psi\in I(\H_\omega)$. Thus one is justified to consider $\im\Phi_\omega$ as $\O_{\K_\omega}$ and assumes that the
function algebra $\im \Phi_\omega$ defines the structure of the $N$-dimensional  complex analytic manifold on $M_\omega$. In such
way we come back to the initial data of the constructions presented in Section \ref{2} and Section \ref{3}, i.e. we reconstruct classical
K\"ahler phase space from the quantum one.

\section*{Acknowledgement}
Author would like to thank T. Goliński and G. Jakimowicz for their interest in the paper. This work was partially supported by KBN grant 2 PO3 A 012 19.


\end{document}